\documentclass[hidelinks,conference]{IEEEtran}

\usepackage{cite}
\usepackage{amsmath,amssymb,amsfonts}
\usepackage{graphicx}
\usepackage{subfigure}
\usepackage{textcomp}
\usepackage[dvipsnames]{xcolor}
\usepackage[acronym,shortcuts]{glossaries}
\usepackage{bm}
\usepackage{caption}
\usepackage{subcaption}
\usepackage{relsize}
\usepackage{soul}
\usepackage[bottom]{footmisc}
\usepackage{algorithm, algpseudocode}
\usepackage{algcompatible}
\usepackage{outlines}
\usepackage{orcidlink}
\usepackage{lipsum,comment}
\usepackage{mathtools}
\usepackage{tabularx,flushend}
\usepackage{amsthm}
\usepackage{dblfloatfix}

\theoremstyle{definition}

\def\BibTeX{{\rm B\kern-.05em{\sc i\kern-.025em b}\kern-.08em
T\kern-.1667em\lower.7ex\hbox{E}\kern-.125emX}}

\newcommand{\trans}[0]{^{\mathsf{T}}}

\newcommand{\herm}[0]{^{\mathsf{H}}}

\newacronym{RPE}{RPE}{radar parameter estimation}
\newacronym{OTFS}{OTFS}{orthogonal time frequency space}
\newacronym{AFDM}{AFDM}{affine frequency division multiplexing}
\newacronym{MIMO}{MIMO}{multiple-input multiple-output}
\newacronym{SISO}{SISO}{single-input single-output}
\newacronym{ISAC}{ISAC}{integrated sensing and communications}
\newacronym{3D}{3D}{three-dimensional}
\newacronym{2D}{2D}{two-dimensional}
\newacronym{1D}{1D}{one-dimensional}
\newacronym{RX}{RX}{receiver}
\newacronym{TX}{TX}{transmitter}
\newacronym{BF}{BF}{beamforming}
\newacronym{mmWave}{mmWave}{millimeter-wave}
\newacronym{SotA}{SotA}{state-of-the-art}
\newacronym{ULA}{ULA}{uniform linear array}
\newacronym{QAM}{QAM}{quadrature amplitude modulation}
\newacronym{ISFFT}{ISFFT}{inverse symplectic finite Fourier transform}
\newacronym{SFFT}{SFFT}{symplectic finite Fourier transform}
\newacronym{AWGN}{AWGN}{additive white Gaussian noise}
\newacronym{OFDM}{OFDM}{orthogonal frequency division multiplexing}
\newacronym{OCDM}{OCDM}{orthogonal chirp division multiplexing}
\newacronym{BS}{BS}{base station}
\newacronym{UE}{UE}{user equipment}
\newacronym{DFT}{DFT}{discrete Fourier transform}
\newacronym{IDFT}{IDFT}{inverse discrete Fourier transform}
\newacronym{TD}{TD}{time-domain}
\newacronym{wlg}{wlg}{without loss of generality}
\newacronym{CP}{CP}{cyclic prefix}
\newacronym{DAFT}{DAFT}{discrete affine Fourier transform}
\newacronym{IDAFT}{IDAFT}{inverse discrete affine Fourier transform}
\newacronym{CPP}{CPP}{\textit{chirp-periodic} prefix}
\newacronym{IDZT}{IDZT}{inverse discrete Zak transform}
\newacronym{DZT}{DZT}{discrete Zak transform}
\newacronym{ICI}{ICI}{inter-carrier interference}
\newacronym{BER}{BER}{bit error rate}
\newacronym{DoF}{DoF}{degrees-of-freedom}
\newacronym{FD}{FD}{full-duplex}
\newacronym{SIMO}{SIMO}{single-input multiple-output}
\newacronym{MISO}{MISO}{multiple-input single-output}
\newacronym{AoD}{AoD}{angle-of-departure}
\newacronym{AoA}{AoA}{angle-of-arrival}
\newacronym{RF}{RF}{radio frequency}
\newacronym{SIM}{SIM}{stacked intelligent metasurfaces}
\newacronym{FIM}{FIM}{flexible intelligent metasurface}
\newacronym{FPGA}{FPGA}{field programmable gate array}
\newacronym{UPA}{UPA}{uniform planar array}
\newacronym{CC}{CC}{communication-centric}
\newacronym{I/O}{I/O}{input-output}
\newacronym{iid}{i.i.d.}{independent and identically distributed}
\newacronym{IoT}{IoT}{internet of things}
\newacronym{V2X}{V2X}{vehicle-to-everything}
\newacronym{NTN}{NTN}{non-terrestrial network}
\newacronym{LEO}{LEO}{low-earth orbit}
\newacronym{THz}{THz}{terahertz}
\newacronym{EM}{EM}{electromagnetic}
\newacronym{RIS}{RIS}{reconfigurable intelligent surface}
\newacronym{DoA}{DoA}{direction-of-arrival}
\newacronym{DD}{DD}{doubly-dispersive}
\newacronym{ODDM}{ODDM}{orthogonal delay-Doppler division multiplexing}
\newacronym{LoS}{LoS}{line-of-sight}
\newacronym{NLoS}{NLoS}{non-line-of-sight}
\newacronym{6G}{6G}{sixth generation}
\newacronym{MPDD}{MPDD}{metasurfaces-parameterized DD}
\newacronym{GaBP}{GaBP}{Gaussian Belief Propagation}
\newacronym{MSE}{MSE}{mean-squared-error}
\newacronym{sIC}{soft IC}{soft interference cancellation}
\newacronym{soft RG}{soft RG}{soft replica generation}
\newacronym{BG}{BG}{belief generation}
\newacronym{SGA}{SGA}{scalar Gaussian approximation}
\newacronym{CLT}{CLT}{central limit theorem}
\newacronym{PDF}{PDF}{probability density function}
\newacronym{QPSK}{QPSK}{quadrature phase-shift keying}
\newacronym{LMMSE}{LMMSE}{linear minimum mean square error}
\newacronym{SNR}{SNR}{signal-to-noise ratio}
\newacronym{QoS}{QoS}{quality of service}
\newacronym{CoV}{CoV}{calculus of variations}
\newacronym{CAPA}{CAPA}{continuous aperture array}
\newacronym{GL}{GL}{Gauss-Legendre}
\newacronym{DDC MIMO}{DDC MIMO}{DD continuous MIMO}
\newacronym{B5G}{B5G}{beyond fifth generation}
\newacronym{VR}{VR}{virtual reality}
\newacronym{XR}{XR}{extended reality}
\newacronym{ITN}{ITN}{intelligent traffic networks}
\newacronym{SAGIN}{SAGIN}{space-air-ground integrated network}
\newacronym{UAV}{UAV}{unmanned aerial vehicle}
\newacronym{MUSIC}{MUSIC}{Multiple Signal Classification}
\newacronym{ICC}{ICC}{integrated communication and computing}
\newacronym{MRC}{MRC}{maximum ratio combining}
\newacronym{LIS}{LIS}{large intelligent surface}

\begin{document}


\title{Low-Complexity Receiver Design for Multicarrier \!CAPA-based Systems in Doubly-Dispersive Channels\!\!}


\author{\IEEEauthorblockN{Kuranage Roche Rayan Ranasinghe$^*$, Giuseppe Thadeu Freitas de Abreu$^*$ and Emil Bj{\"o}rnson$^\dag$}
\IEEEauthorblockA{$^*$\textit{School of Computer Science and Engineering, Constructor University, Bremen, Germany} \\
$^\dag$\textit{School of Electrical Engineering and Computer Science, KTH Royal Institute of Technology, Stockholm, Sweden} \\
Emails: [kranasinghe, gabreu]@constructor.university, emilbjo@kth.se}}




\maketitle

\begin{abstract}
We propose a novel low-complexity receiver design for multicarrier \ac{CAPA} systems operating over \ac{DD} channels.
The receiver leverages a \ac{GaBP}-based framework that hinges only on element-wise scalar operations for the detection of the transmitted symbols.
Simulation results for the \ac{OFDM}, \ac{OTFS}, and \ac{AFDM} waveforms demonstrate significant performance improvements in terms of uncoded \ac{BER} compared to conventional discrete antenna array systems, while maintaining very low computational complexity. 
\end{abstract}

\begin{IEEEkeywords}
Multicarrier, CAPA, GaBP, OFDM, OTFS, AFDM.
\end{IEEEkeywords}

\glsresetall

\vspace{-1ex}
\section{Introduction}


\IEEEPARstart{T}{he} future \ac{6G} of wireless networks are expected to accommodate for numerous distinct applications, such as \ac{ISAC} \cite{LuongCOMMST2025}, massive \ac{IoT} \cite{Chowdhury_6G}, \ac{ICC} \cite{ranasinghe2025flexibledesignframeworkintegrated} and \ac{SAGIN} \cite{CuiCC2022}.
Fulfilling these varied demands necessitates that \ac{6G} networks function in turbulent radio settings, particularly in fast-moving multipath situations like low-earth orbit satellites, vehicle-to-everything scenarios, and unmanned aerial vehicle environments.

Such scenarios naturally lead to so-called \ac{DD} channels \cite{Bliss_Govindasamy_2013}, in which Doppler spreading causes major issues for traditional \ac{OFDM} \cite{wang2006performance}. Consequently, this has spurred the creation of waveforms designed for \ac{DD} channels \cite{RanasingheTWC2025} such as \ac{OTFS} \cite{Hadani_WCNC_2017} and \ac{AFDM} \cite{Bemani_TWC_2023}.

Alongside these developments, another strategy for attaining exceptional network throughput, per-device speed, and detection precision involves the integration of a large number of antennas in a compact area, resulting in the \textit{gigantic} \ac{MIMO} framework \cite{BjornsonOJCOMS2025}.
For this purpose, various cutting-edge array designs have been examined lately, encompassing \acp{LIS} \cite{Hu_TSP_2018}, \acp{SIM} \cite{AnTWC2025}, \acp{FIM} \cite{ranasinghe2025FIM}, and, more recently, \acp{CAPA}.

Due to their natural continuous form, analyzing and enhancing \acp{CAPA} depend on core \ac{EM} principles, which translate the ongoing emitting surface into data-carrying sinusoidal currents spread over the transmitting aperture. These currents produce \ac{EM} fields that are then captured and interpreted by a receiving \ac{CAPA} to extract the intended data.
Although there is abundant research on \ac{BF} strategies for \ac{CAPA} setups (for instance, \cite{WangTWC2025,wang2025beamformingdesigncontinuousaperture,OuyangTWC2025}), low-complexity receiver designs for \ac{DD} channels in \ac{CAPA} systems remain largely unexplored, with the notable exception of \cite{Dardari_JSAC_2020} which, while not deriving a receiver design for a specific system model, offers an information-theoretic perspective on the fundamental limits of receiver architectures with \acp{LIS}.

Building on the channel modeling approach proposed in \cite{ranasinghe2025doublycountinuousmimo}, we fill this research gap by introducing a novel low-complexity receiver design based on the \ac{GaBP} architecture for use in optimized multicarrier \ac{CAPA}-based systems.
Simulation results demonstrate that the proposed receiver achieves significant \ac{BER} improvements compared to conventional discrete systems.

\textit{Notation:} Scalars are denoted by uppercase or lowercase letters, column vectors by bold lowercase letters, and matrices by bold uppercase letters.
The diagonal matrix constructed from vector $\mathbf{a}$ is denoted by $\text{diag}(\mathbf{a})$. For a matrix $\mathbf{A}$, we use $\mathbf{A}\trans$, $\mathbf{A}\herm$, $\mathbf{A}^{1/2}$, and $[\mathbf{A}]_{i,j}$ to represent its transpose, Hermitian, square root, and $(i,j)$-th element, respectively.
The convolution and Kronecker product are denoted by $*$ and $\otimes$, while $\mathbf{I}_N$ and $\mathbf{F}_N$ denote the $N \times N$ identity matrix and the normalized $N$-point \ac{DFT} matrix, respectively.
The sinc function is defined as $\text{sinc}(a) \triangleq \frac{\sin(\pi a)}{\pi a}$, and $j \triangleq \sqrt{-1}$ denotes the imaginary unit.
The Dirac delta function is denoted by $\delta(\cdot)$.
The Lebesgue measure of a Euclidean subspace $\mathcal{S}$ is denoted by $\mu(\mathcal{S})$.
The absolute value and Euclidean norm are denoted by $|\cdot|$ and $\|\cdot\|$, respectively.


\vspace{-2ex}
\section{System, Channel, and Signal Models}
\label{FIM_MIMO_Model}

\begin{figure}[H]
\centering
\includegraphics[width=1\columnwidth]{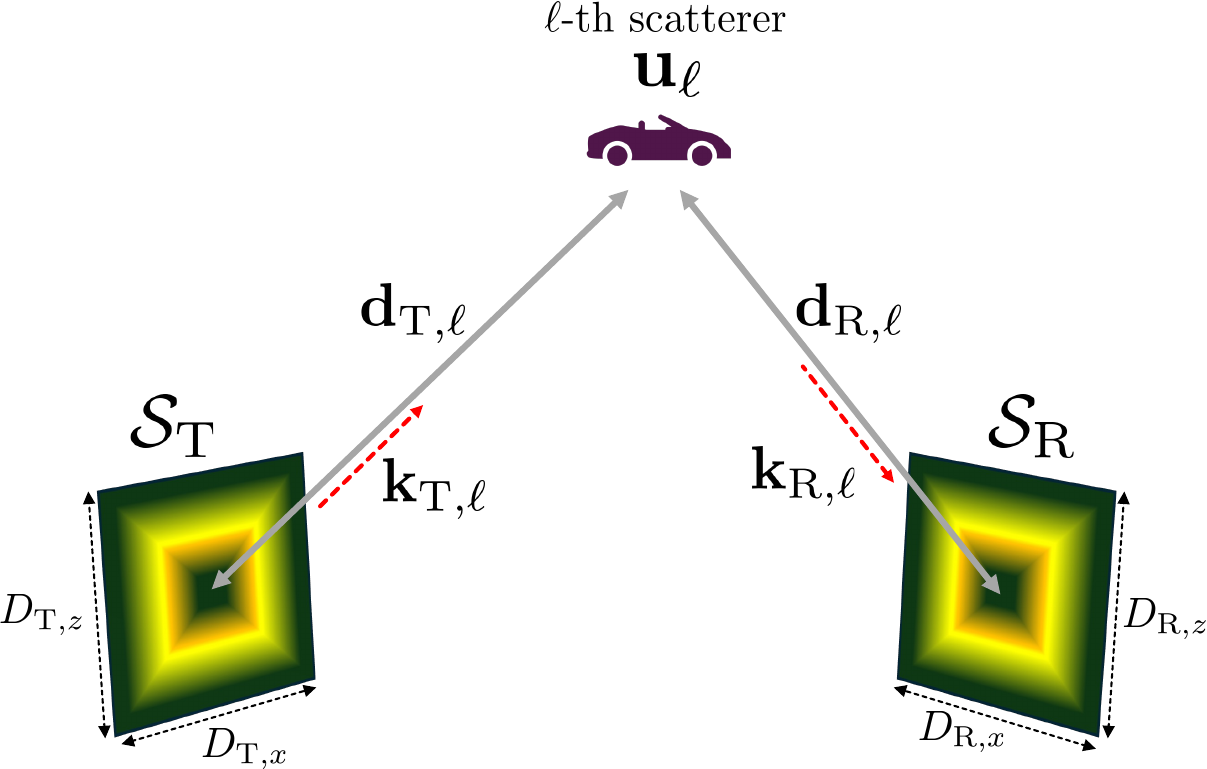}
\caption{Illustration of the point-to-point multicarrier \ac{CAPA}-based system.}
\label{fig:system_model}
\end{figure}

Consider a point-to-point multicarrier system with square-shaped \acp{CAPA} at both the \ac{TX} and \ac{RX}, respectively occupying continuous surfaces $\mathcal{S}_{\mathrm{T}}$ of area $A_\mathrm{T} = \mu(\mathcal{S}_{\mathrm{T}})$ and $\mathcal{S}_{\mathrm{R}}$ of area $A_\mathrm{R} = \mu(\mathcal{S}_{\mathrm{R}})$.
It is assumed that the \ac{TX}-\ac{CAPA} lies on the $x\text{-}z$ plane, centered at the origin, with its two sides parallel to the $x$- and $z$-axes, and side lengths denoted by $D_{\mathrm{T},x}$ and $D_{\mathrm{T},z}$,  respectively, as shown in Fig. \ref{fig:system_model}.

Leveraging typical far-field and wideband assumptions, the \ac{DD} channel between a point $\mathbf{s} \in \mathcal{S}_{\mathrm{T}}$ on the transmit \ac{CAPA} and a point $\mathbf{r} \in \mathcal{S}_{\mathrm{R}}$ on the receive \ac{CAPA} can be expressed via a time-delay impulse response as \cite{ranasinghe2025doublycountinuousmimo}

\vspace{-6ex}
\begin{equation}
\bm{H}(\mathbf{r},\mathbf{s}; \tau, t)
\!=\!\! \sum_{\ell=1}^{L} h_\ell \bm{\Xi}_\ell \delta(\tau\! -\! \tau_\ell) e^{-j 2\pi \nu_\ell t} e^{j \tfrac{2\pi}{\lambda_c} \mathbf{k}_{\mathrm{R},\ell}\trans \mathbf{r}} e^{j \tfrac{2\pi}{\lambda_c} \mathbf{k}_{\mathrm{T},\ell}\trans \mathbf{s}}, 
\label{eq:final_channel_model_H_time-delay}
\end{equation}
where $\tau_\ell$ and $\nu_\ell$ represent the full delay and Doppler shift of the $\ell$-th path, respectively, $\lambda_c$ is the carrier wavelength, $L$ is the total number of paths, and the large-scale path gain is \cite{ranasinghe2025doublycountinuousmimo}
\vspace{-1ex}
\begin{equation}
\label{eq:channel_gain}
\vspace{-1ex}
h_\ell \triangleq \frac{1}{\sqrt{L}}\,\frac{1}{(4\pi)^2 d_{\mathrm{R},\ell}\, d_{\mathrm{T},\ell}},
\end{equation}
where $d_{\mathrm{T},\ell}$ and $d_{\mathrm{R},\ell}$ are the distances from the transmit and receive \acp{CAPA} to the $\ell$-th scatterer, respectively.

In addition, the polarization operator $\bm{\Xi}_\ell \in \mathbb{C}^{3\times 3}$ composes the transverse projectors with the per-path polarization transfer matrix
\vspace{-1ex}
\begin{equation}
\vspace{-1ex}
\label{eq:polarization_matrix}
\bm{\Xi}_\ell \triangleq \big(\mathbf{I}_3 - \mathbf{k}_{\mathrm{R},\ell}\mathbf{k}_{\mathrm{R},\ell}\trans\big)\,
\mathbf{\Gamma}_\ell\,
\big(\mathbf{I}_3 - \mathbf{k}_{\mathrm{T},\ell}\mathbf{k}_{\mathrm{T},\ell}\trans\big),
\end{equation}
where $\mathbf{\Gamma}_\ell \in \mathbb{C}^{3\times 3}$ is the attenuation and polarization transfer matrix of the $\ell$-th path, and $\mathbf{k}_{\mathrm{T},\ell} \in \mathbb{C}^{3\times 1}$ and $\mathbf{k}_{\mathrm{R},\ell} \in \mathbb{C}^{3\times 1}$ are the unit wave vectors at the transmit and receive \acp{CAPA} for the $\ell$-th path, respectively.

\vspace{-1ex}
\subsection{Signal model for DD Waveforms}
\vspace{-1ex}

Let $\mathbf{c}$ be an $NM\times 1$ vector of digitally-modulated information symbols in a constellation $\mathcal{D}$ (with average power $E_\mathrm{C}$) to be transmitted over $N$ subcarriers with $M$ data streams using a \ac{DD} waveform which, for the sake of future convenience, will be assumed to be \ac{AFDM}.\footnote{The distinguishing factors for the models of \ac{OFDM} and \ac{OTFS} will follow with their corresponding numerical results as well for comparison.}
It was shown in \cite{RanasingheTWC2025} that the corresponding received signal, after sampling, rectangular pulse-shaping, propagation through a multipath channel with $L$ paths, each with beamforming matrix $\check{\mathbf{H}}_\ell$, and demodulation, already taking into account the effect of introducing and removing cyclic prefixes, is given by \cite[eqs.~(51) and (52)]{ranasinghe2025doublycountinuousmimo}
\vspace{-1ex}
\begin{equation}
\vspace{-2ex}
\mathbf{y} \!=\! \underbrace{\Bigg(  \sum_{\ell=1}^L \check{\mathbf{H}}_\ell \otimes \overbrace{( \mathbf{\Lambda}_2  \mathbf{F}_{N}  \mathbf{\Lambda}_1 \mathbf{G}_\ell \mathbf{\Lambda}_1\herm  \mathbf{F}_{N}\herm  \mathbf{\Lambda}_2\herm)}^{\mathbf{G}_\ell^\text{AFDM} \in \mathbb{C}^{N \times N}} \Bigg)}_{\bar{\mathbf{H}} \in \mathbb{C}^{NM \times NM}}  \mathbf{c} + \bar{\mathbf{w}} \in \mathbb{C}^{NM\times 1},
\label{eq:DAF_input_output_relation}
\end{equation}
with 
\vspace{-1ex}
\begin{equation}
    \label{eq:check_H_ell}
    \check{\mathbf{H}}_\ell \triangleq \int\limits_{\mathcal{S}_{\mathrm{R}}}\! \int\limits_{\mathrm{S}_{\mathrm{T}}} h_\ell \bm{J}_{\mathrm{R}}\herm(\mathbf{r}) \bm{\Xi}_\ell \bm{J}_{\mathrm{T}}(\mathbf{s}) e^{j \frac{2\pi}{\lambda_c} \mathbf{k}_{\mathrm{R},\ell}\trans \mathbf{r}} e^{j \frac{2\pi}{\lambda_c} \mathbf{k}_{\mathrm{T},\ell}\trans \mathbf{s}} d \mathbf{s} d \mathbf{r},
\end{equation}
where $\mathbf{G}_\ell \triangleq \mathbf{\Phi}_{\ell} \mathbf{Z}^{f_\ell} \mathbf{\Pi}^{\zeta_\ell}$, with $\mathbf{\Phi}_{\ell} \in \mathbb{C}^{N \times N}$, $\mathbf{Z}^{f_\ell} \in \mathbb{C}^{N \times N}$ and $\mathbf{\Pi}^{\zeta_\ell} \in \mathbb{C}^{N \times N}$ being matrices that model\footnote{Due to space limitations, we omit several details such as the sampling rate, the normalization of delays and Doppler shits, etc.} the phase-shift effects of cyclic prefixes, Doppler shifts, and delay spreading, respectively \cite{ranasinghe2025doublycountinuousmimo}.

In \eqref{eq:check_H_ell}, $\bm{J}_{\mathrm{R}}\herm(\mathbf{r}), \bm{J}_{\mathrm{T}}\herm(\mathbf{s}) \in \mathbb{C}^{3 \times M}$ are assumed to represent the continuous beamforming matrices which are computed via an iteration of closed-forms with the approach used in \cite{ranasinghe2025doublycountinuousmimo} to maximize the received power.
It was also shown in \cite{ranasinghe2025doublycountinuousmimo} that similar expressions also result if the information vector $\mathbf{c}$ is transmitted using other waveforms, sufficing to that end to modify the effective channel path matrix $\mathbf{G}_\ell$ accordingly.

In the case of \ac{OFDM} and \ac{OTFS}, instead of the matrix $\mathbf{G}_\ell$ implicitly defined in \eqref{eq:final_channel_model_H_time-delay} we would have respectively
%
\begin{subequations}
\label{eq:Gp_others}
\begin{eqnarray}
\label{eq:Gp_OFDM}
&\mathbf{G}^\text{OFDM}_\ell = \mathbf{F}_N  \mathbf{Z}^{f_\ell} \mathbf{\Pi}^{\zeta_\ell} \mathbf{F}_N\herm,&\\
\label{eq:Gp_OTFS}
&\mathbf{G}^\text{OTFS}_\ell = (\mathbf{F}_{N_1}\otimes\mathbf{I}_{N_2}) \mathbf{Z}^{f_\ell} \mathbf{\Pi}^{\zeta_\ell}  (\mathbf{F}_{N_1}\herm\otimes \mathbf{I}_{N_2}),&
\end{eqnarray}
where we highlight that for \ac{OFDM} and \ac{OTFS} $\mathbf{\Phi}_{\ell}=\mathbf{I}_N$ and the quantities $N_1$ and $N_2$ in the latter equation are both integers such that $N_1 \cdot N_2 = N$.
\end{subequations}

All in all, \eqref{eq:final_channel_model_H_time-delay} and \eqref{eq:check_H_ell}, combined with \eqref{eq:Gp_others} when applicable, fully describe what \ac{AFDM}, \ac{OFDM} and \ac{OTFS} signals are subjected to under a multicarrier \ac{CAPA}-based system.

\section{Proposed GaBP-based Receiver Design}
\label{secReceiver}

From the perspective of receiver design, our objective is to recover the transmitted symbols $\mathbf{c}$, assuming perfect knowledge of the overall channel matrix $\bar{\mathbf{H}} \in \mathbb{C}^{NM \times NM}$. 
To facilitate the development of a \ac{GaBP}-based detector tailored for the multicarrier \ac{CAPA}-based system, we focus on the \ac{I/O} relationship expressed as
\vspace{-1ex}
\begin{equation}
\label{General_I/O_arbitrary}
\mathbf{y} = \bar{\mathbf{H}}  \mathbf{c} + \bar{\mathbf{w}} \in \mathbb{C}^{NM\times 1}.
\end{equation}

Let $\bar{N} \triangleq NM$ and $\bar{M} \triangleq NM$ with the indices $\bar{n} \triangleq \{1,\dots,\bar{N}\}$ and $\bar{m} \triangleq \{1,\dots,\bar{M}\}$.
Then, \eqref{General_I/O_arbitrary} can be expressed element-wise as
\vspace{-1ex}
\begin{equation}
\label{General_I/O_arbitrary_elementwise}
y_{\bar{n}} = \sum_{\bar{m}=1}^{\bar{M}} \bar{h}_{\bar{n},\bar{m}} c_{\bar{m}} + \bar{w}_{\bar{n}}.
\end{equation}

Accordingly, the soft estimate of the $\bar{m}$-th transmitted symbol corresponding to the $\bar{n}$-th received signal $y_{\bar{n}}$, as obtained in the $i$-th iteration of the message-passing algorithm, is represented by $\hat{c}_{\bar{n},\bar{m}}^{(i)}$. 
The associated \ac{MSE} for these estimates at the $i$-th iteration $i$ can then be expressed as
\begin{equation}
\hat{\sigma}^{2(i)}_{c:{\bar{n},\bar{m}}} \triangleq \mathbb{E}_{c} \big[ | c - \hat{c}_{\bar{n},\bar{m}}^{(i-1)} |^2 \big]= E_\mathrm{C} - |\hat{c}_{\bar{n},\bar{m}}^{(i-1)}|^2, \forall (\bar{n},\bar{m}),
\label{eq:MSE_d_k}
\end{equation}
where $\mathbb{E}_{c}$ denotes the expectation over all the possible symbols in the constellation $\mathcal{D}$.

The linear \ac{GaBP} receiver for the relationship portrayed in \eqref{General_I/O_arbitrary_elementwise} typically consists of three major stages as described below.

\subsubsection{Soft Interference Cancellation} 
At the $i$-th iteration, the purpose of the \ac{sIC} stage is to exploit the soft replicas $\hat{c}_{\bar{n},\bar{m}}^{(i-1)}$ obtained from the preceding iteration, in order to generate the data-oriented \ac{sIC} signals $\tilde{y}_{c:\bar{n},\bar{m}}^{(i)}$.
\newpage

Exploiting \eqref{General_I/O_arbitrary_elementwise}, the \ac{sIC} signals are given as
\vspace{-1ex}
\begin{align}
\label{eq:d_soft_IC}
\vspace{-1ex}
\tilde{y}_{c:\bar{n},\bar{m}}^{(i)} &= y_{\bar{n}} - \sum_{e \neq \bar{m}} \bar{h}_{\bar{n},e} \hat{c}_{\bar{n},e}^{(i)}, \\[-1ex]
&= \bar{h}_{\bar{n},\bar{m}} c_{\bar{m}} + \underbrace{\sum_{e \neq \bar{m}} \bar{h}_{\bar{n},e}(c_e - \hat{c}_{\bar{n},e}^{(i)}) + \bar{w}_{\bar{n}}}_\text{interference + noise term}.\nonumber
\end{align}

By applying the \ac{SGA}, the interference and noise components in the previous expression can be approximated as complex Gaussian distributed. 
Consequently, the conditional \acp{PDF} of the \ac{sIC} signals become
\vspace{-1ex}
\begin{equation}
\label{eq:cond_PDF_d}
\vspace{-1ex}
\!\!p_{\tilde{\mathrm{y}}_{\mathrm{c}:\bar{n},\bar{m}}^{(i)} \mid \mathrm{c}_{\bar{m}}}(\tilde{y}_{c:\bar{n},\bar{m}}^{(i)}|c_{\bar{m}}) \propto \mathrm{exp}\bigg[ -\frac{|\tilde{y}_{c:\bar{n},\bar{m}}^{(i)}\! -\! \bar{h}_{\bar{n},\bar{m}} c_{\bar{m}}|^2}{\tilde{\sigma}_{c:\bar{n},\bar{m}}^{2(i)}} \bigg]\!,
\end{equation}
with their conditional variances given by
\vspace{-1ex}
\begin{equation}
\label{eq:soft_IC_var_d}
\vspace{-1ex}
\tilde{\sigma}_{c:\bar{n},\bar{m}}^{2(i)} = \sum_{e \neq \bar{m}} \left|\bar{h}_{\bar{n},e}\right|^2 \hat{\sigma}^{2(i)}_{c:{\bar{n},e}} + \sigma^2_w.
\end{equation}


\subsubsection{Belief Generation}

Within the belief generation stage, the \ac{SGA} is applied under the assumptions that $\bar{N}$ is sufficiently large and that the estimation errors in $\hat{c}_{\bar{n},\bar{m}}^{(i-1)}$ are mutually independent. 
This allows for the construction of initial symbol estimates ($i.e.,$ beliefs) for all the data symbols.

As a direct result of the \ac{SGA}, and using the conditional \acp{PDF} in equation \eqref{eq:cond_PDF_d}, the extrinsic \acp{PDF} can be obtained as
\vspace{-1ex}
\begin{equation}
\label{eq:extrinsic_PDF_d}
\vspace{-1ex}
\prod_{e \neq \bar{n}} p_{\tilde{\mathrm{y}}_{\mathrm{c}:e,\bar{m}}^{(i)} \mid \mathrm{c}_{\bar{m}}}(\tilde{y}_{c:e,\bar{m}}^{(i)}|c_{\bar{m}}) \propto \mathrm{exp}\bigg[ - \frac{(c_{\bar{m}} - \bar{c}_{\bar{n},\bar{m}}^{(i)})^2}{\bar{\sigma}_{c:\bar{n},\bar{m}}^{2(i)}} \bigg],
\end{equation}
where the extrinsic means and variances are
\vspace{-1ex}
\begin{equation}
\label{eq:extrinsic_mean_d}
\bar{c}_{\bar{n},\bar{m}}^{(i)} = \bar{\sigma}_{c:\bar{n},\bar{m}}^{(i)} \sum_{e \neq \bar{n}} \frac{\bar{h}^*_{e,\bar{m}} \tilde{y}_{c:e,\bar{m}}^{(i)}}{ \tilde{\sigma}_{c:e,\bar{m}}^{2(i)}},
\end{equation}
\vspace{-1ex}
\begin{equation}
\vspace{-1ex}
\label{eq:extrinsic_var_d}
\bar{\sigma}_{c:\bar{n},\bar{m}}^{2(i)} = \bigg( \sum_{e \neq \bar{n}} \frac{|\bar{h}_{e,\bar{m}}|^2}{\tilde{\sigma}_{c:e,\bar{m}}^{2(i)}} \bigg)^{\!\!\!-1},
\end{equation}
with $\bar{h}^*_{e,\bar{m}}$ denoting the complex conjugate of $\bar{h}_{e,\bar{m}}$.

\subsubsection{Soft Replica Generation}

In the final stage of soft replica generation, the previously obtained beliefs are denoised using a Bayes-optimal criterion, thereby yielding the final update for the estimates of the target variables. 
For \ac{QPSK} modulation\footnote{We adopt \ac{QPSK} for simplicity; however, \ac{wlg}, denoisers for alternative modulation formats can also be constructed when necessary.}, the Bayes-optimal denoiser is expressed as
\vspace{-1ex}
\begin{equation}
\vspace{-1ex}
\hat{c}_{\bar{n},\bar{m}}^{(i)}\! =\! q_x \bigg(\! \text{tanh}\!\bigg[ 2c_d \frac{\Re{\bar{c}_{\bar{n},\bar{m}}^{(i)}}}{\bar{\sigma}_{c:{\bar{n},\bar{m}}}^{2(i)}} \bigg]\!\! +\! j \text{tanh}\!\bigg[ 2c_d \frac{\Im{\bar{c}_{\bar{n},\bar{m}}^{(i)}}}{\bar{\sigma}_{{c}:{\bar{n},\bar{m}}}^{2(i)}} \bigg]\!\bigg),\!\!
\label{eq:QPSK_denoiser}
\end{equation}
where $q_x \triangleq \sqrt{E_\mathrm{C}/2}$ denotes the amplitude of the real and imaginary components of the chosen \ac{QPSK} symbols, and the corresponding variance is updated according to \eqref{eq:MSE_d_k}.

Once $\hat{c}_{\bar{n},\bar{m}}^{(i)}$ is obtained from  \eqref{eq:QPSK_denoiser}, the final outputs are refined through a damping process, which mitigates the risk of convergence to local minima caused by erroneous hard-decision replicas. 
Introducing a damping factor $0 < \beta_c < 1$ gives
\vspace{-1ex}
\begin{equation}
\vspace{-1ex}
\label{eq:d_damped}
\hat{c}_{\bar{n},\bar{m}}^{(i)} = \beta_c \hat{c}_{\bar{n},\bar{m}}^{(i)} + (1 - \beta_c) \hat{c}_{\bar{n},\bar{m}}^{(i-1)}.
\end{equation}

\vspace*{-3ex}
\begin{algorithm}[H]
\caption{GaBP-based Detection for \ac{CAPA}-based Systems}
\label{alg:proposed_decoder}
\setlength{\baselineskip}{11pt}
\textbf{Input:} received signal $\mathbf{y}\in\mathbb{C}^{\bar{N}\times 1}$, complete channel matrix $\bar{\mathbf{H}} \in \mathbb{C}^{\bar{N}\times \bar{M}}$, no. of \ac{GaBP} iterations $i_{G}$, data constellation power $E_\mathrm{C}$, noise variance $\sigma^2_w$ and damping factor $\beta_c$. \\
\textbf{Output:} $\hat{\mathbf{c}}$ 
\vspace{-2ex} 
\begin{algorithmic}[1]  
\STATEx \hspace{-3.5ex}\hrulefill
\STATEx \hspace{-3.5ex}\textbf{Initialization}
\STATEx \hspace{-3.5ex} - Set iteration counter to $i=0$ and amplitudes $q_x = \sqrt{E_\mathrm{C}/2}$.
\STATEx \hspace{-3.5ex} - Set initial data estimates to $\hat{c}_{\bar{n},\bar{m}}^{(0)} = 0$ and corresponding 
\STATEx \hspace{-2ex} variances to $\hat{\sigma}^{2(0)}_{c:{\bar{n},\bar{m}}} = E_\mathrm{C}, \forall \bar{n},\bar{m}$.
\STATEx \hspace{-3.5ex}\hrulefill
\STATEx \hspace{-3.5ex}\textbf{for} $i=1$ to $i_\text{G}$ \textbf{do}: $\forall \bar{n}, \bar{m}$
\STATE Compute \ac{sIC} data signal $\tilde{y}_{c:{\bar{n},\bar{m}}}^{(i)}$ and its corresponding variance $\tilde{\sigma}^{2(i)}_{c:{\bar{n},\bar{m}}}$ from equations \eqref{eq:d_soft_IC} and \eqref{eq:soft_IC_var_d}.
\STATE Compute extrinsic data signal belief $\bar{c}_{\bar{n},\bar{m}}^{(i)}$ and its corresponding variance $\bar{\sigma}_{c:{\bar{n},\bar{m}}}^{2(i)}$ from equations \eqref{eq:extrinsic_mean_d} and \eqref{eq:extrinsic_var_d}.
\STATE Compute denoised and damped data signal estimate $\hat{c}_{\bar{n},\bar{m}}^{(i)}$ from equations \eqref{eq:QPSK_denoiser} and \eqref{eq:d_damped}.
\STATE Compute denoised and damped data signal variance $\hat{\sigma}_{c:{\bar{n},\bar{m}}}^{2(i)}$ from equations \eqref{eq:MSE_d_k} and \eqref{eq:MSE_d_m_damped}.

\STATEx \hspace{-3.5ex}\textbf{end for}
\STATE Calculate $\hat{c}_{\bar{m}}, \forall \bar{m}$ (equivalently $\hat{\mathbf{c}}$) using equation \eqref{eq:d_hat_final_est}. 

\end{algorithmic}
\end{algorithm}
\vspace{-2ex}

Similarly, the variances $\hat{\sigma}^{2(i)}_{c:{\bar{n},\bar{m}}}$ are initially updated via \eqref{eq:MSE_d_k} and then damped using
\vspace{-1ex}
\begin{equation}
  \vspace{-1ex}
\label{eq:MSE_d_m_damped}
\hat{\sigma}^{2(i)}_{c:{\bar{n},\bar{m}}} = \beta_c \hat{\sigma}_{c:{\bar{n},\bar{m}}}^{2(i)} + (1-\beta_c) \hat{\sigma}_{c:{\bar{n},\bar{m}}}^{2(i-1)}.
\end{equation}

Finally, the consensus update of the estimates can be obtained via
\vspace{-1ex}
\begin{equation}
  \vspace{-1ex}
\label{eq:d_hat_final_est}
\hat{c}_{\bar{m}} = \bigg( \sum_{\bar{n}=1}^{\bar{N}} \frac{|\bar{h}_{\bar{n},\bar{m}}|^2}{\tilde{\sigma}_{c:\bar{n},\bar{m}}^{2(i_\text{max})}} \bigg)^{\!\!\!-1} \! \! \bigg( \sum_{\bar{n}=1}^{\bar{N}} \frac{\bar{h}^*_{\bar{n},\bar{m}} \tilde{y}_{c:\bar{n},\bar{m}}^{(i_\text{max})}}{ \tilde{\sigma}_{c:\bar{n},\bar{m}}^{2(i_\text{max})}} \bigg).
\end{equation}

The complete pseudocode for the proposed detection procedure is summarized above in Algorithm \ref{alg:proposed_decoder}.

\section{Performance Analysis}
\label{secSymResults}

\vspace{-0.5ex}
\subsection{Complexity Analysis}

The computational complexity of the proposed \ac{GaBP} detection algorithm is dictated by the number of element-wise operations, and its per-iteration computational complexity is given by $\mathcal{O}(\bar{N}\bar{M})$. 
Notice that this complexity is much less than that of typical detection methods such as the \ac{LMMSE}, which is $\mathcal{O}(\bar{N}^3)$ due to the large matrix inversion involved.

\vspace{-0.5ex}
\subsection{Numerical Results}

Unless otherwise specified, the system parameters listed in Table \ref{tab:simulation_parameters} are persistently used throughout this section.
For the optimization in \cite{ranasinghe2025doublycountinuousmimo}, the maximum iterations $i_{\mathrm{S}}$ were set to 20 and the \ac{GL} quadrature sample points were set to $\bar{M}_x = \bar{M}_z = \bar{K}_x = \bar{K}_z = 10$.
For performance comparisons, we mainly focus on following \cite{WangTWC2025,SanguinettiTWC2023}, where we can define conventional spatially discrete antenna arrays, with the continuous surfaces $\mathcal{S}_\mathrm{T}$ and $\mathcal{S}_\mathrm{R}$ occupied with discrete antennas with effective aperture $A_d = \frac{\lambda^2_c}{4\pi}$ and antenna spacing $d_t = d_r = \frac{\lambda}{2}$.
Further information on the discrete optimization to be solved can be found in \cite{ranasinghe2025doublycountinuousmimo}.
For the \ac{GaBP}, $i_{G}$ was set to 20 and $\beta_c$ set to 0.5, with $E_\mathrm{C}$ set to 1.
\newpage

\begin{figure}[H]
\subfigure[{\footnotesize $N=64$}]%
{\includegraphics[width=\columnwidth]{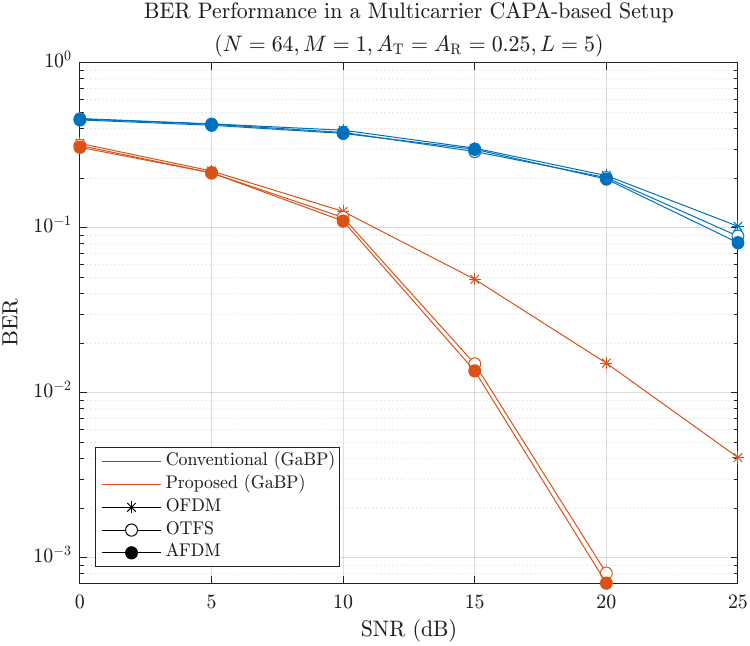}
\label{fig:BER_N64}}
\vspace{-2ex}\\
\vspace{-1ex}
\subfigure[{\footnotesize $N=144$}]%
{\includegraphics[width=\columnwidth]{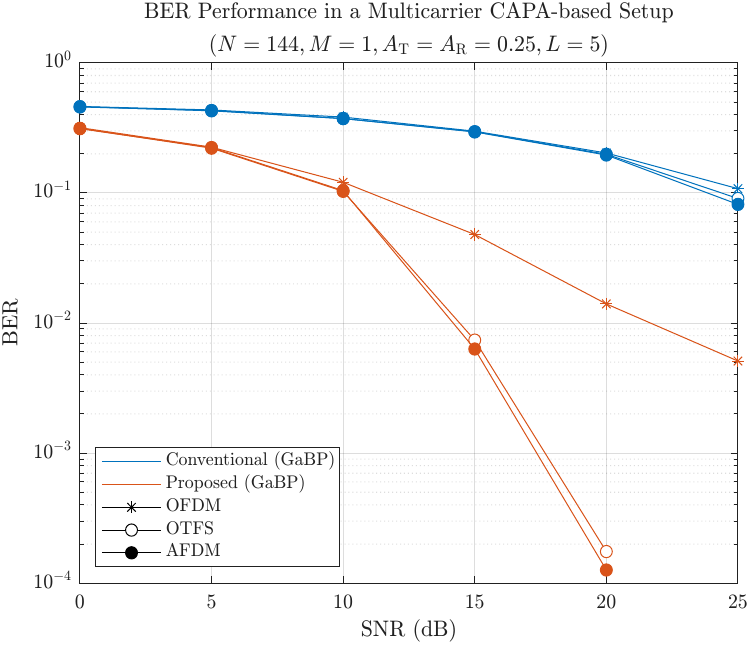}
\label{fig:BER_N144}}
\vspace{-1ex}
\caption{Comparison of \ac{BER} performances of the continuous and conventional discrete schemes.}
\label{fig:BER_MSE_Alg2}
\end{figure}

Fig.~\ref{fig:BER_MSE_Alg2} compares the uncoded \ac{BER} of the \ac{GaBP}-based receiver for both the continuous and conventional discrete settings, showing a significant improvement achieved by continuous apertures when path loss is taken into account.
\vspace{-2ex}
\begin{table}[H]
\centering
\caption{System Parameters}
\label{tab:simulation_parameters}
\begin{tabular}{|c|c|c|}
\hline
\textbf{Parameter} & \textbf{Symbol} & \textbf{Value} \\
\hline
Carrier Frequency & $f_c$ & 2.4 GHz \\
\hline
Carrier Wavelength & $\lambda_c$ & 0.1249 m \\
\hline
Bandwidth and Sampling Frequency & $B$, $F_S$ & 1 MHz \\
\hline
Number of Subcarriers & $N$ & 64, 144 \\
\hline
Total RF chains & $M$ & 1 \\
\hline
Number of Channel Scatterers & $L$ & 5 \\
\hline
Maximum Range and Velocity & $R_\text{max}$, $V_\text{max}$ & 1500 m, 122 m/s \\
\hline
Aperture Size & $A_\mathrm{T}, A_\mathrm{R}$ & 0.25 m$^2$ \\
\hline
\end{tabular}
\end{table}
\vspace{-5ex}

In addition, \ac{OTFS} and \ac{AFDM} significantly outperform the classical \ac{OFDM} waveform in terms of \ac{BER}, which is expected due to the high mobility conditions leading to the degradation of the \ac{OFDM} subcarrier structure.

\section{Conclusion}

This work introduces a low-complexity receiver architecture tailored for multicarrier \ac{CAPA} systems in \ac{DD} channel environments.
The design employs a \ac{GaBP}-inspired approach that relies solely on simple scalar computations per element to decode the sent data symbols.
Numerical evaluations across \ac{OFDM}, \ac{OTFS}, and \ac{AFDM} modulations reveal substantial \ac{BER} enhancements over traditional discrete setups, all achieved with minimal processing demands.

\bibliographystyle{IEEEtran}
\bibliography{references}

\end{document}